\documentclass[num-refs]{wiley-article}

\usepackage{siunitx}

\papertype{Original Article}
\paperfield{Journal Section}

\title{Characterizing normal perinatal development of the human brain structural connectivity}

\author[1]{Yihan Wu}
\author[2]{Lana Vasung}
\author[1]{Camilo Calixto}
\author[1]{Ali Gholipour}
\author[1]{Davood Karimi}

\affil[1]{Computational Radiology Laboratory (CRL), Department of Radiology, Boston Children's Hospital, and Harvard Medical School, USA}

\affil[2]{Department of Pediatrics at Boston Children's Hospital, and Harvard Medical School, Boston, Massachusetts, USA.}

\corraddress{Davood Karimi, Department of Radiology, Boston Children's Hospital, 55 Fruit Street, 2ndfloor Main Building, Boston, MA 02215, USA.}
\corremail{davood.karimi@childrens.harvard.edu}

\fundinginfo{National Institutes of Health (NIH): R01HD110772, R01NS128281, R01NS106030, R01EB031849, R01EB032366, R01HD109395. National Science Foundation (NSF): 212306.}

\runningauthor{Wu et al.}

\begin{document}

\begin{frontmatter}
\maketitle

\begin{abstract}

Early brain development is characterized by the formation of a highly organized structural connectome. The interconnected nature of this connectome underlies the brain's cognitive abilities and influences its response to diseases and environmental factors. Hence, quantitative assessment of structural connectivity in the perinatal stage is useful for studying normal and abnormal neurodevelopment. However, estimation of the connectome from diffusion MRI data involves complex computations. For the perinatal period, these computations are further challenged by the rapid brain development and imaging difficulties. Combined with high inter-subject variability, these factors make it difficult to chart the normal development of the structural connectome. As a result, there is a lack of reliable normative baselines of structural connectivity metrics at this critical stage in brain development. In this study, we developed a computational framework, based on spatio-temporal averaging, for determining such baselines. We used this framework to analyze the structural connectivity between 33 and 44 postmenstrual weeks using data from 166 subjects. Our results unveiled clear and strong trends in the development of structural connectivity in perinatal stage. Connection weighting based on fractional anisotropy and neurite density produced the most consistent results. We observed increases in global and local efficiency, a decrease in characteristic path length, and widespread strengthening of the connections within and across brain lobes and hemispheres. We also observed asymmetry patterns that were consistent between different connection weighting approaches. The new computational method and results are useful for assessing normal and abnormal development of the structural connectome early in life.

\end{abstract}

\end{frontmatter}

\section{Introduction}
Understanding the structural connectivity of the brain is a central goal of neuroscience. The interconnected nature of the structural connectome is one of the most intrinsic and most important properties of the brain \cite{collin2013ontogeny, rossini2019methods, fornito2015connectomics, ingalhalikar2014sex}. Structural connectomics underlies our cognitive abilities, influences the progression and spread of neuropathologies, and shapes the brain's response to injury \cite{scheinost2017does, schmithorst2018structural, jakab2015disrupted}. There is mounting evidence that the metrics computed from a graph representation of the structural connectome are strong markers of various disorders such as attention-deficit hyperactivity disorder \cite{cao2013probabilistic, hong2014connectomic}, schizophrenia \cite{wheeler2014review, goldsmith2018update}, multiple sclerosis \cite{sbardella2013dti, fleischer2017increased}, and aging \cite{zhao2015age, wen2011discrete}.

Diffusion-weighted magnetic resonance imaging (dMRI) is a unique tool for quantitative assessment of structural brain connectivity \cite{shi2017connectome, tymofiyeva2014structural}. Orientation of white matter tracts can be estimated in each imaging voxel from the dMRI signal. Tractography methods can then be applied to trace virtual streamlines that represent these tracts across the brain volume. The white matter tracts connect different parts of the brain's gray matter, which can naturally be treated as nodes of a graph. The streamlines will represent the edges connecting these nodes. The strength of each edge can be quantified in terms of the number of streamlines or based on measures of white matter micro-structure integrity. Graph-theoretic metrics can be employed to quantitatively characterize the structural connectome and to perform cross-subject comparisons \cite{rubinov2010complex, bullmore2011brain}. Reliable and reproducible computation of the structural connectome based on the dMRI measurements is challenging, and the computed metrics should be interpreted with caution. Nonetheless, much effort has been dedicated to improving the reproducibility of quantitative connectivity analysis with dMRI \cite{zhang2022quantitative, zalesky2010whole, sotiropoulos2019building, yeh2016correction}. Constant technical advancements have enhanced the accuracy and reliability of this approach \cite{qi2015influence, sotiropoulos2019building}. Furthermore, there have been great strides in improving our understanding of the potentials and limitations of this technique \cite{fornito2013graph, jones2010challenges}. Consequently, dMRI-based quantitative assessment of structural brain connectivity is increasingly utilized to study brain development, maturation, aging and degeneration \cite{griffa2013structural, assaf2008diffusion}.

The great majority of prior works have focused on pediatric and adult brains. Comparatively, much less is known about the structural brain networks very early in life. Image acquisition difficulties, scarcity of postmortem material for deriving the gold standard, and a lack of reliable quantitative analysis methods have made it difficult to study the structural connectivity at the perinatal stage. Furthermore, because of methodological variations in computing the structural connectome, inherent limitations of dMRI, and high inter-subject variability, it has been difficult to establish normative references for longitudinal and population studies. This represents a critical gap in knowledge as it is well known that adult-like topological structures and a highly structured brain connectome develop very early in life \cite{van2015neonatal, marami2017temporal, silbereis2016cellular, kostovic2006development, ouyang2019delineation, bayer2005human}. 

Because early brain development is vulnerable to diseases and environmental factors, quantitative structural connectivity analysis can have immensely important clinical and scientific applications. For example, prenatal exposure to maternal stress, which affects 10-35\% of children \cite{maselko2015effect, sawyer2010pre, kinney2008prenatal}, may disrupt the development of brain connectome \cite{scheinost2017does, chen2016toward, bronson2014prenatal}. These disruptions can have long-lasting impacts on neurodevelopment and they are a risk factor for disorders such as autism spectrum disorder, attention deficit hyperactivity disorder (ADHD), depression, and schizophrenia \cite{kinney2008prenatal, chen2016toward, constantinof2016programming, khashan2008higher, brown2012epidemiologic, li2010attention, hoerder2015development}. Another example is congenital heart disease (CHD), which is the most common birth defect and is suspected to alter the brain connectome in neonates \cite{schmithorst2018structural}. Moreover, it has been shown that the structural connectivity analysis has a unique potential for characterizing, classifying, and understanding the clinical heterogeneity of various brain malformations, such as the agenesis of the corpus callosum, that begin in the fetal period \cite{jakab2015disrupted, meoded2011prenatal, kasprian2013assessing, millischer2022feasibility}. Hence, there is an urgent need for methods and resources to enable accurate and reproducible quantitative assessment of structural brain connectivity in the perinatal stage. Such methods can significantly enhance our understanding of perinatal brain development and improve our knowledge of the neurodevelopmental processes that shape the structure and function of the brain for the rest of life.

The goal of this work is to develop a new methodology for assessing the normal development of the brain's structural connectivity in the perinatal stage. The proposed approach is based on spatial alignment (also referred to as spatial normalization) and averaging of data from subjects of the same age. To achieve precise spatial alignment of white matter structures between subjects, we perform the registrations based on maps of diffusion tensor and fiber orientation distribution. This strategy reduces the impact of inter-subject variability and low data quality, thereby amplifying and highlighting the main trends in structural connectivity that take place due to brain development. As a consequence, we expect that this approach should be able to reconstruct normative structural connectivity metrics that can serve as references for reliable assessment and comparison of normal and abnormal brain development at this critical stage. We demonstrate the effectiveness of this approach by applying it to a large cohort of subjects scanned between 33 and 44 weeks of postmenstrual age (PMA) and analyzing several important metrics of structural connectivity.

\section{MATERIALS AND METHODS}

\subsection{Data}
\label{ssec:Subjects}

We used the MRI data from the second release of the Developing Human Connectome Project (dHCP) study \cite{bastiani2019automated}. All subjects included in this analysis were healthy, i.e., without major brain focal lesions or any clinically significant abnormalities based on expert evaluation of structural MRI. To conduct our analysis, we considered postmenstrual ages (PMAs) between 33 and 44 weeks. This period is characterized by interhemispheric synchronisation and a gradual resolution of subplate that corresponds to  establishing permanent brain circuitry \cite{kostovic2010development}. The PMA encompasses the summation of gestational age in weeks (interval between the initial day of the last menstrual period and the day of delivery) and chronological age in weeks, which signifies the time elapsed since birth. The PMA was rounded to the nearest week. For PMA of 35 weeks, for example, we used subjects scanned between 34.5 and 35.5 postmenstrual weeks. For PMAs around 38 weeks, the dHCP dataset contained many more subjects than needed for our analysis. Our recent work as well as works of other researchers have shown minimal or no changes in the quality of spatio-temporal atlases when more than 15 subjects are used in each age group \cite{karimi2022atlas, pietsch2019framework}. Hence, we used at most 15 subjects for each PMA. For the earliest age of 33 weeks only seven subjects were available, but that was still sufficient for our analysis.

Structural MRI data for each subject included T1w and T2w multi-slice fast spin-echo images acquired with an in-plane resolution of 0.8mm, slice thickness of 1.6mm, and slice overlap of 0.8mm in axial and sagittal directions. The dMRI data was collected with a set of spherically optimized directions at four b-values: $b=0$ ($n=20$), $b=400$ ($n=64$), $b=1000$ ($n=88$), and $b=2600$ ($n=128$). Pre-processing of raw structural data included bias correction with the N4 algorithm, motion corrected volumetric reconstructions of the multi-slice acquisitions, brain extraction using BET from the FSL software package \cite{jenkinson2005bet2}, and tissue segmentation using the DRAW-EM algorithm developed for neonatal brains \cite{makropoulos2014automatic, makropoulos2018developing}. A fetal neuroanatomist (L.V.) carefully inspected and verified the segmentations. Pre-processing of the raw dMRI data included susceptibility-induced distortion correction, correction of eddy current-induced distortions and subject motion, followed by super-resolution volume reconstruction and registration to structural images. The details of the pre-processing operations are described in \cite{bastiani2019automated} and \cite{makropoulos2018developing}. We further applied denoising on the dMRI data and resampled all dMRI and anatomical data (i.e., T2 images and tissue segmentations) to an isotropic resolution of 1mm.

\subsection{Computational pipeline}
\label{ssec:ComputationalPipeline}

\begin{figure*}[!htb]
    \centering
    \includegraphics[width=1.0\linewidth]{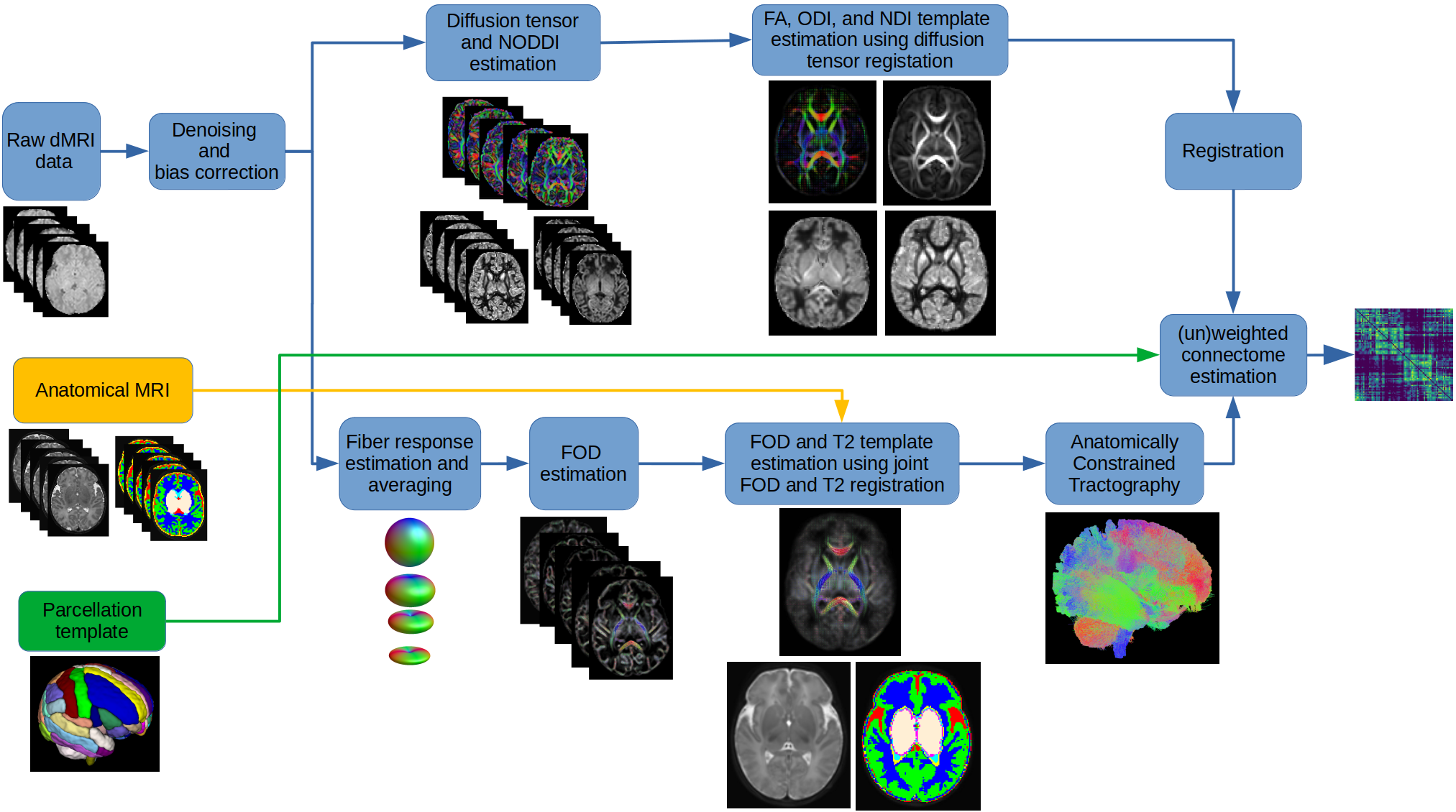}
    \caption{The proposed computational pipeline for computing population-averaged age-specific structural connectomes.}
\label{fig:schematic}
\end{figure*}

Figure \ref{fig:schematic} shows the data processing pipeline for computing population-averaged age-specific connectomes. The pipeline has two main branches. One branch uses fiber orientation distribution (FOD)-based registration to compute a tractogram for each age. The other branch uses diffusion tensor-based registration to compute maps of micro-structural biomarkers. The FOD-based alignment could have been used to also compute atlases of micro-structural biomarkers. However, we found that a diffusion tensor-based registration resulted in sharper and more detailed micro-structural biomarker maps. Different steps of the pipeline are described below. Note that this pipeline is applied separately for each age group to compute a separate structural connectome for every week between 33 and 44 postmenstrual weeks.

\subsubsection{Computing age-specific FOD templates and tractograms}
\label{sssec:fodpipeline}

We used the multi-shell multi-tissue constrained spherical deconvolution (MSMT-CSD) \cite{jeurissen2014multi} for FOD estimation. This method is based on deconvolving the dMRI signal with signature response functions from white matter, gray matter, and cerebrospinal fluid. We first estimated these response functions separately for each subject in an age group and then created an average response function for that age. The average response function was used to estimate the FOD images for each subject in the age group. A white matter FOD template was then estimated using symmetric diffeomorphic registration of the white matter FOD maps of all subjects in the age group using the method of \cite{raffelt2011symmetric}. The deformations computed based on the white matter FODs were also used to warp the T2 images and tissue segmentation maps. Voxel-wise averaging and majority voting, respectively, were used to estimate a T2 template and a tissue segmentation template for that age. Anatomically-constrained tractography \cite{smith2012anatomically} with a probabilistic streamline tracing method \cite{tournier2010improved} was then applied using the FOD and tissue segmentation templates. We empirically set the maximum angle between successive streamline tracing steps to 30 degrees and the FOD amplitude cut-off threshold of 0.01 as the stopping criterion. We randomly seeded all voxels in the brain volume and generated a total of five million valid streamlines.

\subsubsection{Computing age-specific templates of tissue micro-structure biomarkers}
\label{sssec:dtipipeline}

There is no consensus on the proper weighting of the edges in a structural connectome. It is possible to compute the edge weight/strength values based on tractography data alone, for example in terms of the streamline count. However, there is growing evidence for superiority of utilizing biomarkers of tissue micro-structure integrity to weight the connections \cite{qi2015influence, yeh2021mapping}. In this work, we used biomarkers derived from the diffusion tensor and the Neurite Orientation Dispersion and Density Imaging (NODDI) models \cite{zhang2012noddi}. 

We estimated the diffusion tensor with the iterative weighted least squares method of \cite{veraart2013weighted} using the measurements in the $b=1000$ shell. We computed the fractional anisotropy (FA) image from the diffusion tensor image. We fitted the NODDI-Watson model to the full multi-shell data and computed the Orientation Dispersion Index (ODI) and the Neurite Density Index (NDI) \cite{zhang2012noddi}. As suggested by \cite{guerrero2019optimizing}, we lowered the initial value of parallel diffusivity from $1.7 \times 10^{-9} m^2/s$ to $1.4 \times 10^{-9} m^2/s$ in order to better fit neonatal brain data.

Subsequently, we computed a template for these biomarkers using nonlinear diffusion tensor-based alignment algorithm of \cite{zhang2006deformable} implemented in the DTI-TK software package. These templates were then registered to the T2 template map for the same age group using affine registration. Note that the T2 and FOD templates were co-registered by design, as shown in Figure \ref{fig:schematic}. Hence, after being registered to the T2 template, these biomarker templates could be used to weight the streamlines computed based on the FOD template.

\subsubsection{Computing the structural connectome}
\label{sssec:connectome_computation}

To define the connectome nodes, we used the Edinburgh Neonatal (ENA50) Atlas \cite{blesa2020peak}. This atlas included 107 regions of interest (ROIs) from 53 structures with bilateral representation in addition to the corpus callosum. We only utilized the cortical grey matter parcellations, subcortical grey matter structures, and cerebellar parcellations, resulting in a total of 98 nodes. After excluding the white matter structures and ventricles, we registered this parcellation to our computed age-specific templates using deformable registration of the T2 image from the ENA50 atlas to the T2 template estimated by our pipeline. Using the gray matter parcellations as the graph nodes and streamlines as the edges, we computed the structural connectomes. Afterwards, the Spherical-deconvolution Informed Filtering of Tractograms 2 (SIFT2) algorithm \cite{smith2015sift2} was applied to compute cross-sectional area multipliers to ensure the streamline densities reflected the density of the underlying white matter fibers. Additionally, we computed the mean of microstructural biomarkers along streamlines connecting each pair of nodes to obtain $w_{\text{FA}}(i,j)$, $w_{\text{NDI}}(i,j)$, and $w_{1-\text{ODI}}(i,j)$, which were then used to weight the connections. The negative sign for ODI is standard practice and it is because ODI is a measure of fiber dispersion, whereas we should assign larger weights to higher microstructural integrity.

The procedure described above follows the state of the art approach for computing the structural connectome \cite{smith2020quantitative}. The results presented in this paper mostly follow this analysis. Nonetheless, we also present and discuss the connectivity results after applying a normalization operation proposed by \cite{batalle2017early}. This normalization aims to ensure that different connectomes are equal in terms of the total network strength. It normalizes each connectome as $w_{nX}(i,j) = w_{X}(i,j) / \sum_{\forall i,j} w_{X}(i,j)$, where X refers to the connectome weighting (SIFT2, FA, NDI, or 1-ODI). The rationale behind this normalization strategy is that it facilitates comparison of the connectomes in terms of network topology by reducing the influence of total network strength. In other words, it is anticipated that this normalization will equalize the total network strength for all connectomes, thereby making the graph metrics independent of the total network strength. As a result, it is expected that this normalization would enhance the capability of the computed connectome to describe the topological and organizational properties of the brain \cite{batalle2017early}.

\subsubsection{Computing the connectivity metrics}
\label{sssec:connectivity_metrics}

After computing the connectome as described above, we computed five standard and widely-used structural connectivity measures: characteristic path length (CPL), global efficiency (GE), local efficiency (LE), clustering coefficient (CC), and small-worldness index (SWI). CPL and GE are measures of network integration, which quantifies brain's ability to incorporate information across distant brain regions \cite{rubinov2010network}. CC and LE are measures of network segregation, which reflects the capability for specialized processing to occur within interconnected groups of brain regions \cite{rubinov2010network}. SWI is a measure of network topology. All connectivity measures were computed using the Brain Connectivity Toolbox \cite{rubinov2010complex}.

\subsubsection{Age regression}
\label{sssec:age_regression}

We used general linear models (GLMs) to estimate the effect of age on the structural brain connectivity measures using the network measures as dependent variables and PMA as the independent variable. We used the R statistics package to perform all GLM analyses. We considered a $p$ value of less than 0.05 to be significant.

\subsubsection{Edge-wise association with age}
\label{sssec:edge_wise_association}

We assessed the correlations between the individual connections and PMA to characterize the changes in connectivity weights with age. This edge-wise association analysis was performed on un-normalized as well as normalized SIFT2-, FA-, ND-, (1-ODI)-weighted connections. For this analysis, we only considered the connections that were common to all ages. We used the Spearman's rank correlation coefficient ($\rho$) to quantify the association between edge-wise connection strengths with PMA. We used the Bonferroni correction to account for multiple comparisons in order to control the family-wise error rate at 0.05. Furthermore, we merged all the nodes in each lobe and computed the connectome among the lobes. We assessed the correlation between the PMA and connections between the lobes. Similar to the node-wise analysis, we only considered the connections that were common to all PMAs. We computed the Spearman's rank correlation coefficient to assess the association between lobe-wise connections and PMA. We applied the False Discovery Rate (FDR) correction to control the family-wise error rate at 0.05.

\subsubsection{Asymmetry in brain connectivity}
\label{sssec:Asymmetry}

In order to analyze the differences between the structural connectivity in the left and right brain hemispheres, we computed the laterality index $\text{LI} = \frac{\text{Right}-\text{Left}}{\text{Right}+\text{Left}}$ for the SIFT2-, FA-, ND-, and (1-ODI)-weighted connections. LI \textless 0 indicates leftward asymmetry, whereas LI \textgreater 0 indicates rightward asymmetry. We performed linear regression analyses to assess the effect of age on the asymmetry of the brain connectivity.

\section{RESULTS}
\label{sec:results}

\begin{figure*}[!htb]
    \centering
    \includegraphics[width=1.0\linewidth]{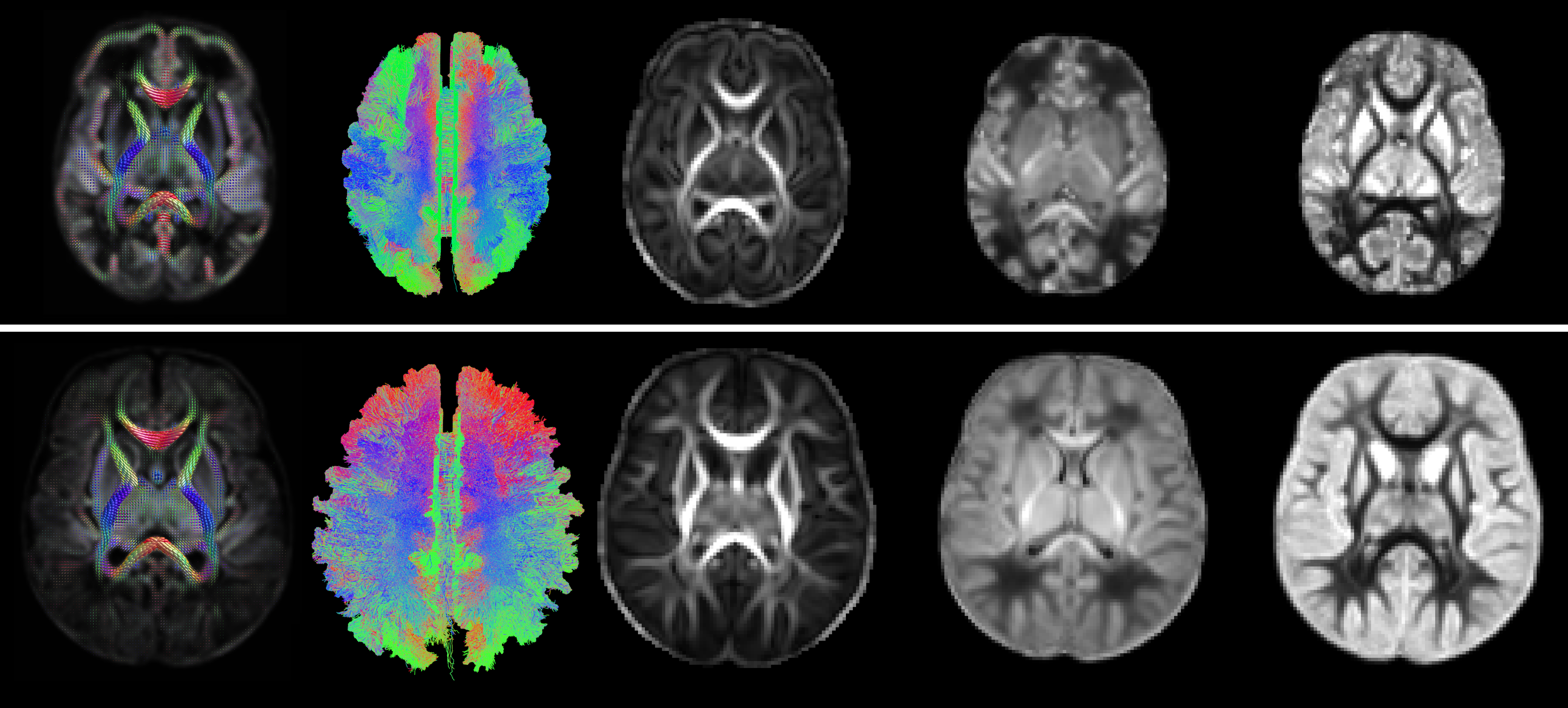}
    \caption{From left to right: example FOD atlas, tractogram, FA, ND, and ODI atlases generated by our computational pipeline for 35 weeks (top) and 43 weeks (bottom).}
\label{fig:atlas}
\end{figure*}

\subsection{Spatio-temporal atlases}
\label{ssec:atlases}

Figure \ref{fig:atlas} shows selected views of the atlases reconstructed by our computational pipeline for PMAs of 35 and 43 weeks. The atlases portray a detailed representation of the brain's structure. To ensure accuracy, an expert visually assessed the atlases to confirm that they were free from errors and artifacts, verify the orientation of the FODs, and determine the correctness of the computed tractograms.

\subsection{Association between connectivity metrics and PMA}
\label{ssec:measures}

\begin{figure*}[!htb]
    \centering
    \includegraphics[width=1.0\linewidth]{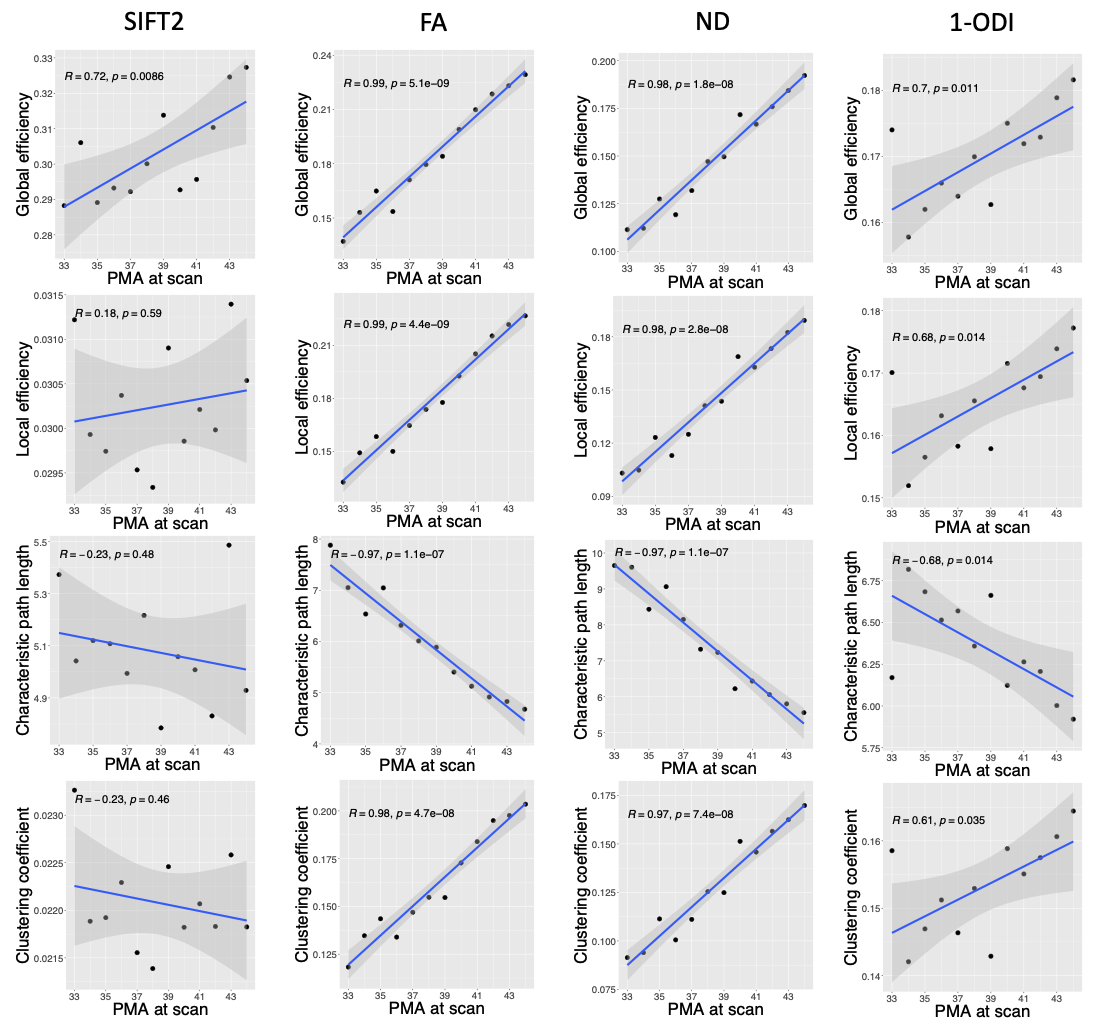}
    \caption{Plots of different structural connectivity measures versus PMA for the connectome edge weighting based on SIFT2, FA, NDI, and 1-ODI.}
\label{fig:measures_unnormalized}
\end{figure*}

\begin{figure*}[!htb]
    \centering
    \includegraphics[width=1.0\linewidth]{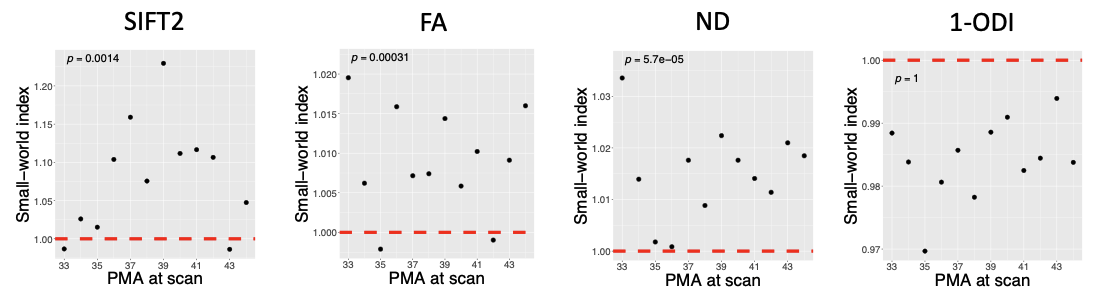}
    \caption{Plots of small-worldness index versus PMA for connectome weighting based on SIFT2, FA, NDI, and 1-ODI (one-sample one-tailed t-test).}
\label{fig:swi}
\end{figure*}

The observed trends in brain connectivity measures as a function of PMA are presented in Figure \ref{fig:measures_unnormalized}. Overall, they depict an increase in global efficiency (GE), local efficiency (LE), and clustering coefficient (CC) and a decrease in characteristic path length (CPL). For the SIFT2-weighted connectome, these trends are not very strong ($|R| \in [0.18, 0.72]$). For ODI-weighted connectome, linear correlations are stronger ($|R| \in [0.61, 0.70]$). For connectomes weighted by FA and ND, on the other hand, there is unmistakable strong linear correlations with $|R| \geq 0.97$. GE is positively correlated with PMA in connectomes weighted by SIFT2 (R = 0.72, p = 0.009), FA (R = 0.99, p \textless 0.001), NDI (R = 0.98, p \textless 0.001), and 1-ODI (R = 0.70, p = 0.011). Similarly, LE is positively correlated with PMA in connectomes weighted by FA (R = 0.99, p \textless 0.001), NDI (R = 0.98, p \textless 0.001), and 1-ODI (R = 0.68, p = 0.014). CPL is negatively correlated with PMA in connectomes weighted by FA (R = -0.97, p \textless 0.001), NDI (R = -0.97, p \textless 0.001), and 1-ODI (R = -0.68, p = 0.014). For CC, on the other hand, the results are not consistent between the SIFT2-weighted connectome and the other connectomes. While CC is negatively correlated with PMA in the connectome weighted by SIFT2 (not statistically significant), it is positively correlated with PMA in connectomes weighted by FA, NDI, and 1-ODI.

Figure \ref{fig:swi} shows the small-worldness index (SWI) values computed with different connectome weighting schemes. For SWI, instead of the correlation with PMI, we are interested in knowing whether the computed values are larger than one, which would indicate small-world network properties. We used one-sample t-tests to test the hypothesis that SWI was significantly larger than one. As shown in the figure, this test shows that SWI for the connectome weighting based on SIFT2, FA, and NDI is significantly larger than one, whereas for the connectome weighted based on 1-ODI all SWI values were smaller than one.

\subsection{Node-wise associations with age}
\label{ssec:age_association}

\begin{figure*}[!htb]
    \centering
    \includegraphics[width=0.8\linewidth]{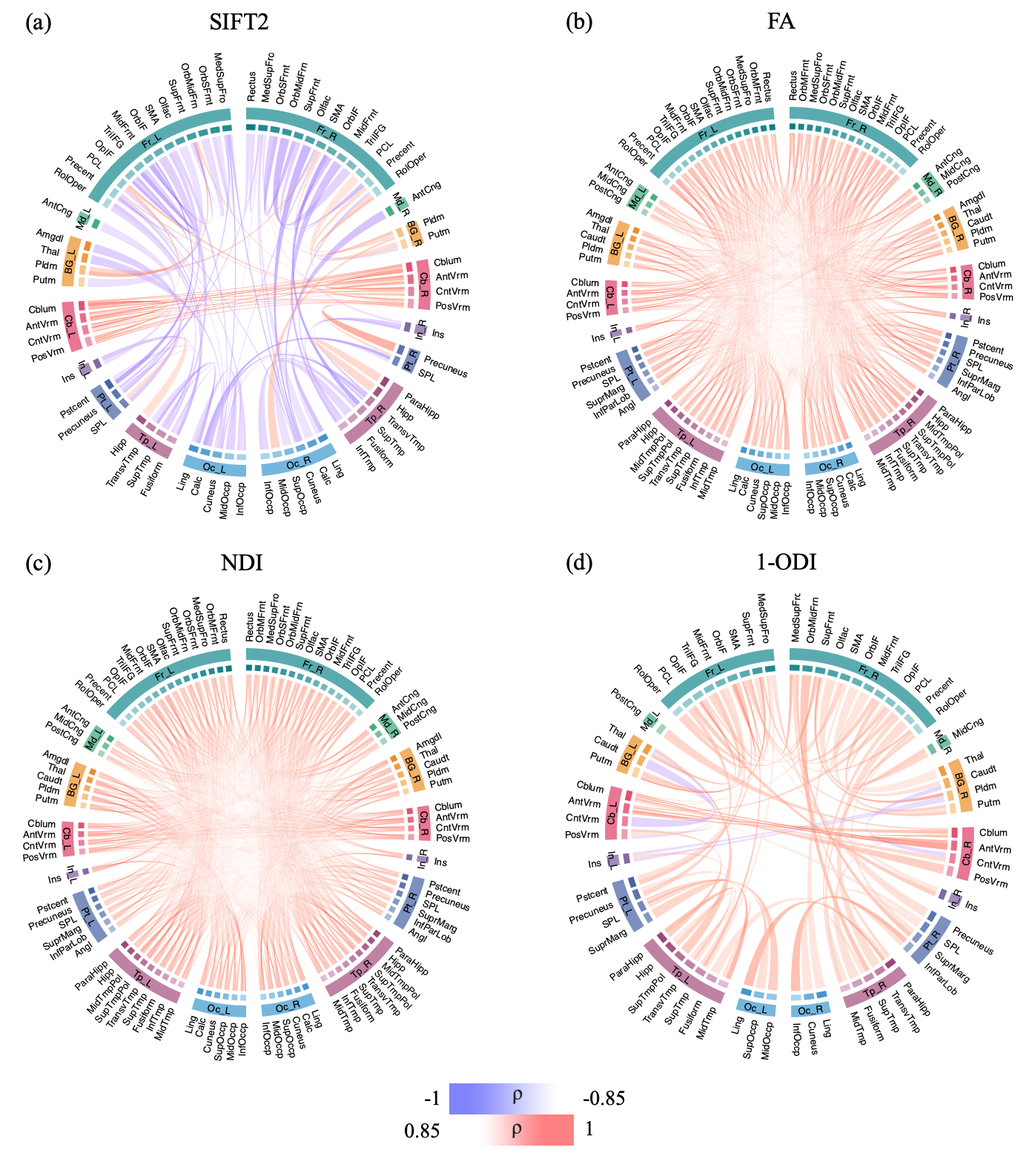}
    \caption{These figures show the connectome edges that are significantly correlated with PMA, quantified in terms of Spearman's rank correlation coefficient ($\rho$) after Bonferroni correction. The color intensity and thickness of the edges are proportional to $\rho$.}
\label{fig:edge_association}
\end{figure*}

Figure \ref{fig:edge_association} shows the association between connection strength and PMA for different connection weighting schemes. For the SIFT2-weighted connectome, where no measure of microstructural integrity of the white matter are included, some of the connections become stronger over time, in particular the intra-cerebellar connections. The connections between the remaining nodes mostly become weaker with increasing PMA. In particular, connections within each of the two hemispheres become weaker, including the connections within the frontal lobe and connections between frontal lobe and occipital and temporal lobes. The connectomes that have been computed by including measures of white matter integrity (FA, NDI, and 1-ODI), on the other hand, paint an entirely different picture. All connections become stronger with PMA, with the exception of a few sporadic connections in the connectome weighted based on 1-ODI. These results show a consistent increase in the strength of the connections that is primarily driven by improved neural density, reduced dispersion, and increased myelination.

\begin{figure*}[!htb]
    \centering
    \includegraphics[width=0.8\linewidth]{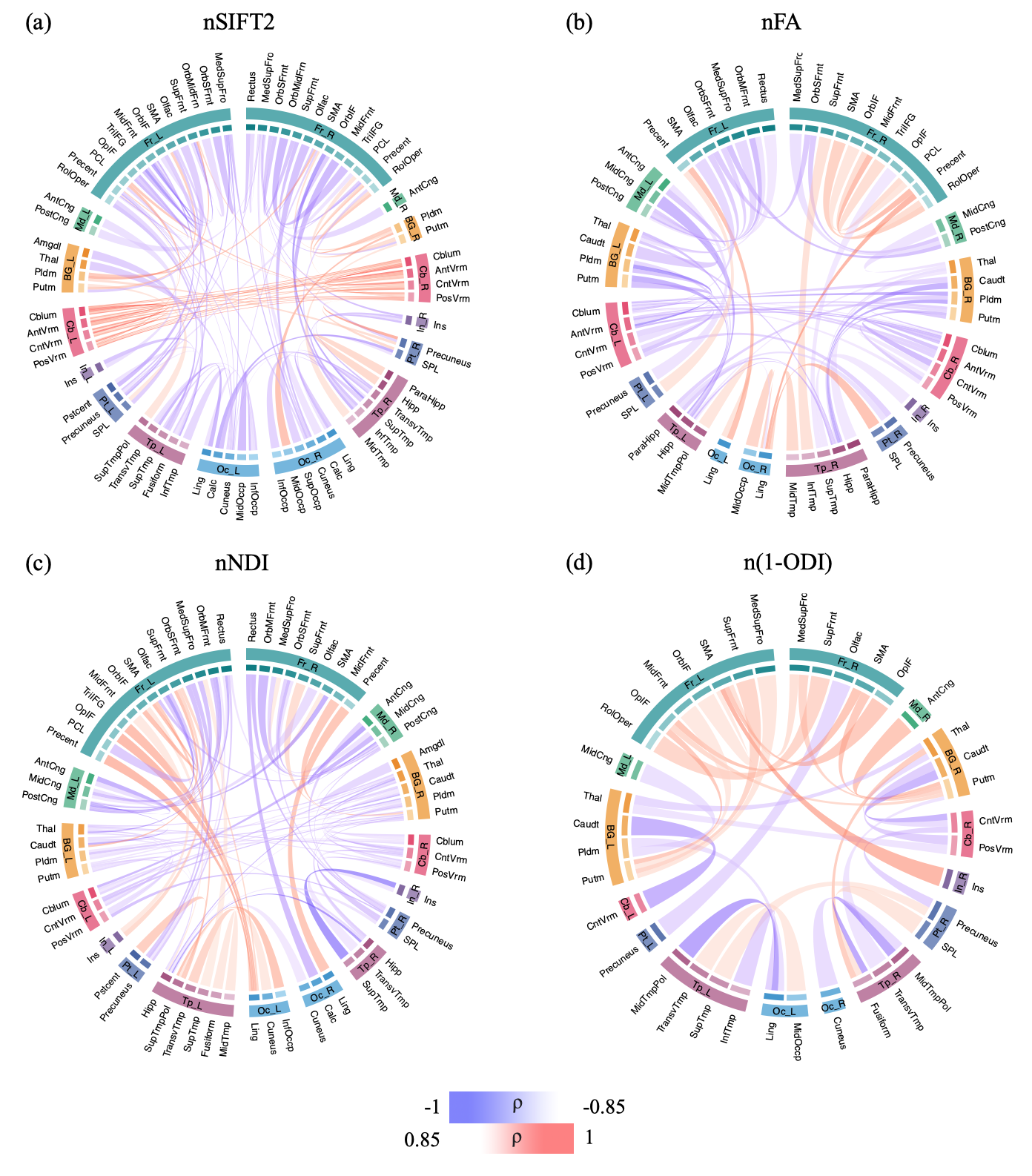}
    \caption{The association between connection strength and PMA in the connectomes normalized by the total connectome strength. The association is quantified in terms of Spearman's rank correlation coefficient ($\rho$) after Bonferroni correction. The color intensity and thickness of the edges are proportional to $\rho$.}
\label{fig:normalized_edge_association}
\end{figure*}

Figure \ref{fig:normalized_edge_association} shows the correlation between PMA and connection strengths in the connectomes that are normalized in terms of the total network strength. Compared with the connectomes shown in Figure \ref{fig:edge_association}, the normalized connectomes display a more complex picture with significant regional variations in maturation. The normalized FA-weighted connectome, for instance, shows an increase in strength for some connections, including within the right frontal lobe, between occipital and temporal lobes, and between precentral and lingual. On the other hand, the connection strength decreases for several connections, including connections within the cerebellum, between BG and cerebellum, between frontal lobe and posterior cingulate, between medial lobe and BG and hippocampus, as well as inter-hemisphere connections. The normalized NDI-weighted connectome shows more increasing connection strengths in the left than in the right hemisphere. Specifically, connection strengths increase between the frontal lobe, insula, temporal, and occipital lobes in the left hemisphere. In the right hemisphere, a few connections become stronger between the frontal lobe and other regions. Inter-hemispheric connections, on the other hand, show much slower rates of change. The normalized (1-ODI)-weighted connectome shows a strengthening of the connections within the frontal lobe and between frontal lobe and putamen, but a weakening of connections between BG and cerebellum and between occipital and temporal lobes.

\subsection{Correlation between PMA and connections between the lobes}
\label{ssec:lobe_association}

\begin{figure*}[!htb]
    \centering
    \includegraphics[width=1\linewidth]{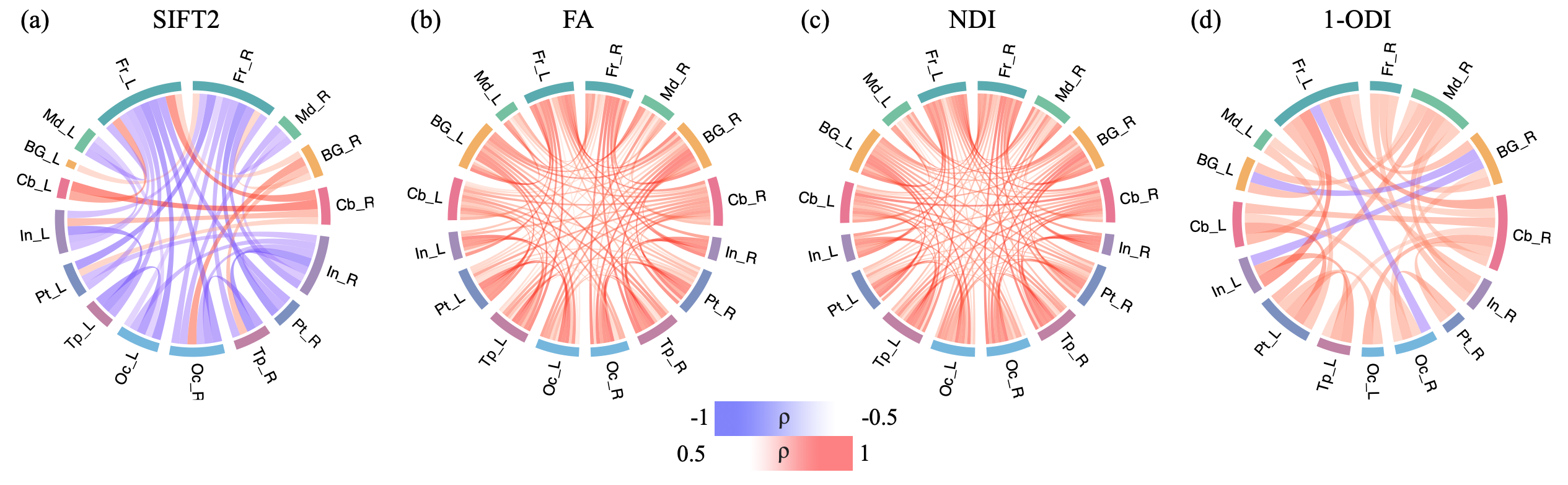}
    \caption{Lobe-wise connections that are significantly correlated with PMA. These connectomes show the Spearman's rank correlation coefficient ($\rho$) after FDR correction. The color intensity and thickness of the edges are proportional to $\rho$.}
\label{fig:age_association_cluster}
\end{figure*}

As shown in Figure \ref{fig:age_association_cluster}, we observe some strong positive correlations with PMA for several connections in the SIFT2-weighted connectome, including connections between cerebellum, BG and the rest of the brain. However, this connectome shows many more negative correlations with PMA, predominantly between the frontal lobes and the other lobes such as temporal and occipital lobes in the same hemisphere as well as the other hemisphere. The connectomes weighted with FA and NDI, on the other hand, display a near-uniform increase in the connection strength with PMA. The (1-ODI)-weighted connectome shows an overall similar pattern of increasing connection strength with a few sporadic decreasing connection strengths.

\begin{figure*}[!htb]
    \centering
    \includegraphics[width=1\linewidth]{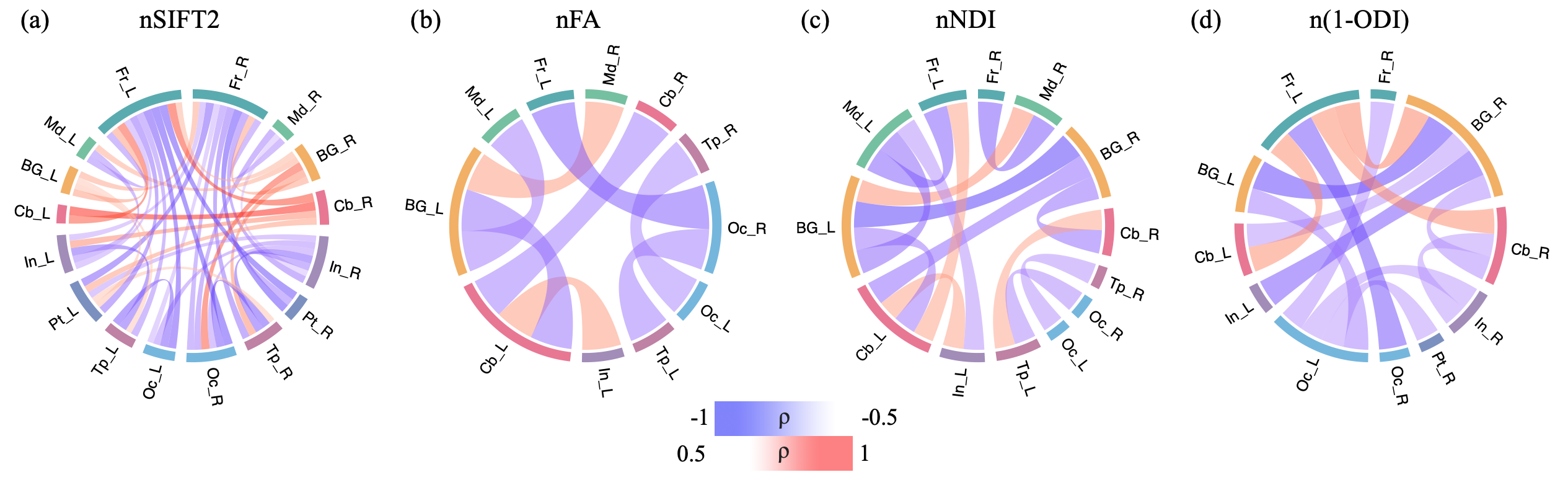}
    \caption{Lobe-wise connections that are significantly correlated with PMA in the connectomes that have been normalized in terms of the total network strength. These connectomes show the Spearman's rank correlation coefficient ($\rho$) after FDR correction. The color intensity and thickness of the edges are proportional to $\rho$.}
\label{fig:normalized_age_association_cluster}
\end{figure*}

Figure \ref{fig:normalized_age_association_cluster} shows the association between lobe-wise connection strengths and PMA in the connectomes that have been normalized in terms of the total network strength. Similar to Figure \ref{fig:normalized_edge_association}, and unlike Figure \ref{fig:age_association_cluster}, the relations displayed in this figure are complex and do not lend themselves to a simple description. Both FA- and NDI-weighted connectomes display weakening connections between multiple lobes within and across the two brain hemispheres, although a few of the connections become stronger such as the connection between the insula and cerebellum in the left hemisphere. The SIFT2-weighted connectome shows a similar overall pattern, but with many more connections.

\subsection{Brain asymmetry}
\label{ssec:brain_asymmetry}

\begin{figure*}[!htb]
    \centering
    \includegraphics[width=1.0\linewidth]{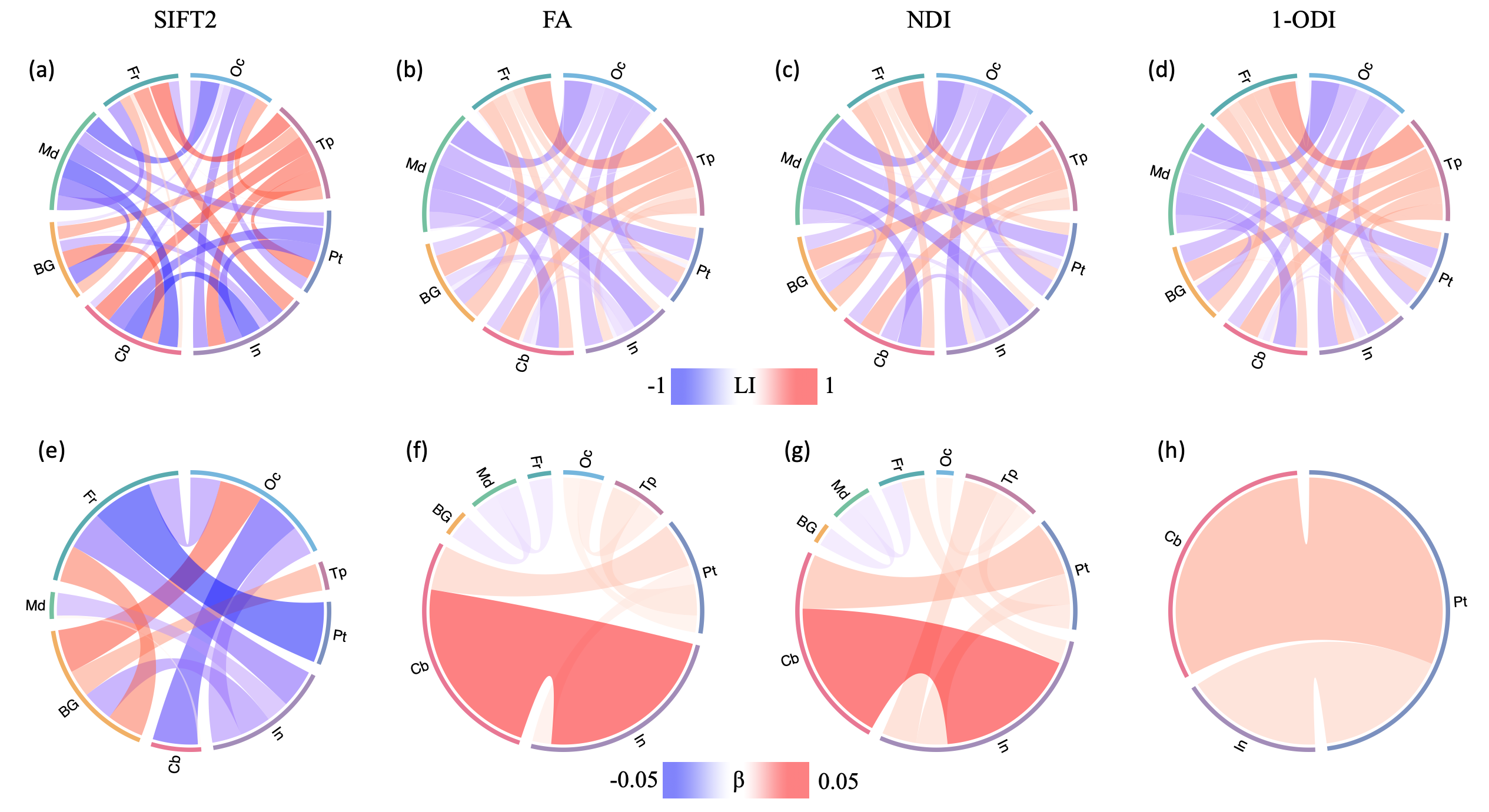}
    \caption{Top: Brain asymmetry quantified in terms of the laterality index (LI). The connections with leftward asymmetry are colored in blue, while the connections with right asymmetry are in red. The color intensity and thickness of the links are proportional to the LI value. Bottom: Correlation between LI and PMA. The color intensity and thickness of the links are proportional to the regression slope $\beta$. A blue color indicates an increasing leftward asymmetry with PMA, while red indicates increasing rightward asymmetry.}
\label{fig:Asymmetry}
\end{figure*}

The results of asymmetry analysis are presented in Figure \ref{fig:Asymmetry}. The asymmetry patterns revealed by the SIFT2-, FA-, NDI-, and (1-ODI)-weighted connectomes, although complex, display remarkable similarities (Figure \ref{fig:Asymmetry}(a)-((d)). The connections between the frontal and temporal lobes and the other lobes show right-ward asymmetry, while the connections between medial and occipital lobes and the other lobes display left-ward asymmetry. The SIFT2-weighted connectome shows asymmetry patterns that are largely similar to the FA-, NDI-, and (1-ODI)-weighted connectomes, but overall it shows stronger asymmetry compared with the other three connectomes.

Figures \ref{fig:Asymmetry}(e)-(h) show the correlation between connection asymmetry and PMA. The patterns for FA- and NDI-weighted connectomes are largely similar, but markedly different than the SIFT2- and (1-ODI)-weighted connectomes. Connections among occipital, temporal, parietal lobe, insula, and cerebellum show an increase in right-ward asymmetry, while connections among BG, medial, and frontal lobes show slight increases in left-ward asymmetry. The sharpest changes occur between cerebellum and insula in the FA- and NDI-weighted connectomes. The associations between LI and PMA in the SIFT2-weighted connectome show quite different patterns. More of the connections display significant changes in asymmetry with PMA. For example, the connection between frontal and parietal lobes shows a strong increase in left-ward asymmetry, which is not observed in the other three connectomes.

\section{DISCUSSION}

In this work, we proposed a computational framework for quantitative assessment of the development of structural connectome in the perinatal stage. We computed the structural connectome and connectivity metrics using the SIFT2 algorithm and micro-structural biomarkers including FA, NDI, and 1-ODI for connection weighting. Our results showed that the proposed framework could unveil strong relationships between several important measures of brain connectivity and PMA. Our analysis of the correlation between connection strength and PMA showed a consistent and wide-spread increase in node-wise and lobe-wise connection strengths in the connectomes weighted by FA and NDI. 

Several prior studies have examined the development of brain structural connectivity from infancy to adolescence using graph theoretical approaches \cite{bhroin2020reduced, hagmann2010white, ouyang2022flattened, baker2015developmental}. There have also been a few studies on the structural connectivity in preterm neonates \cite{batalle2017early, van2015neonatal, brown2014structural, ratnarajah2013structural}. However, all prior works have been conducted on individual subjects. To the best of our knowledge, our work is the first to develop a methodology to assess the development of the structural connectome in the perinatal stage using spatio-temporal normalization and averaging. Therefore, our results provide new insights about the normal development of the structural connectome in this critical period.

Some of the observations in this study are consistent with the findings of prior works. Increases in local efficiency and global efficiency \cite{batalle2017early, bhroin2020reduced, song2017human} and a decrease in characteristic path length \cite{van2015neonatal, brown2014structural} with PMA have been reported in prior studies. The increase in global efficiency and decrease in characteristic path length indicate an increase in network integration, which translates into improved ability of the brain to integrate information from distant regions of the brain and enhanced efficiency of communication between those regions \cite{rubinov2010complex, sporns2016networks}. The increase in local efficiency, on the other hand, indicates an increase in network segregation, which means an increased ability of the brain to support specialized information processing by interconnected clusters of brain regions \cite{rubinov2010complex, sporns2016networks}. Some other of our findings do not agree with those reported in the literature. For example, it has been reported that SWI increases significantly with age in the perinatal stage \cite{batalle2017early, van2015neonatal, brown2014structural}. Our analysis did not reveal such a trend, as shown in Figure \ref{fig:swi}. Some studies have also found that the clustering coefficient increases with PMA (e.g., \cite{brown2014structural}). This is similar to our results with the connectomes weighted by FA, NDI, and 1-ODI, but different from our results with the SIFT2-weighted connectome, as shown in Figure \ref{fig:measures_unnormalized}.

Although some prior works have reported qualitatively similar results, there are important quantitative differences. Most importantly, some of the trends discovered in our work are very strong and show much less variability than those observed in prior works \cite{van2015neonatal, batalle2017early, bhroin2020reduced}. Batalle et al. \cite{batalle2017early}, for example, found significant positive correlations between local and global efficiency and PMA in connectomes weighted by FA and NDI. However, their computed Spearman's correlation coefficients ranged between 0.638 and 0.713. Bhroin et al. reported Spearman's correlation coefficients between 0.343 and 0.400 \cite{bhroin2020reduced}. For characteristic path length, Van Den Heuvel et al. reported $R=0.83$ \cite{van2015neonatal}. In our results, the change in global efficiency, local efficiency, and characteristic path length in the connectomes weighted by FA and NDI follow strong linear trends between 33 and 44 weeks, with $|R|$ very close to one. We attribute this to the effectiveness of our proposed computational method in averaging the data from multiple subjects to reduce the inter-subject variability. Because of the very low variability, these plots can be used as normative references for studying normal and abnormal brain development in the perinatal stage.

Some prior works have also analyzed the change in connection strengths with PMA. Batalle et al. observed significant increases in average FA-weighted and NDI-weighted connection strengths \cite{batalle2017early}. In terms of individual connections, they observed significant increases in many connections in the connectomes weighted with NDI and FA and \emph{decreases} in the connectome weighted by 1-ODI. Their results with FA- and (1-ODI)-weighting included several connections with opposite changes. Therefore, there are important differences between some of those results and our findings, which show uniform increases in terms of FA and NDI and near-uniform increases in terms of 1-ODI (Figure \ref{fig:edge_association}). Furthermore, our observed correlations are much stronger, with many connections showing a Spearman's correlation coefficient $|\rho| \in [0.85,1.0]$ whereas in \cite{batalle2017early} the strongest correlations show $|\rho| \in [0.65,0.80]$. There are also many qualitative differences between our results and those of \cite{batalle2017early}. For example, in the normalized NDI-weighted connectome, we observed increasing connection strengths between the frontal and occipital lobes and decreasing connection strengths between the frontal and medial lobes in both hemispheres (Figure \ref{fig:normalized_edge_association}). No such changes were observed by \cite{batalle2017early}. Brown et al. assessed the change in the connection strength in terms of streamline count and FA with PMA \cite{brown2014structural}. Their study included 47 subjects between 27 and 45 weeks PMA, with 23 of the subjects scanned twice. They observed that in terms of streamline count and FA, respectively, 664 and 1009 of the connections changed significantly with PMA and that most of these changes had a positive slope. For FA-weighted connectome, 83\% of those 1009 connections significantly increased in strength while 17\% of them showed a significant decrease. As shown in Figure \ref{fig:edge_association}, in our results all significantly changing connection strengths have a positive slope. It is possible that this is due to the suppression of unreliable and noisy results by our framework, which may be impossible to achieve when data from individual subjects are considered.

Our analysis of the correlation between connection strength and PMA showed a consistent and wide-spread increase in node-wise and lobe-wise connection strength in the connectomes weighted by FA and NDI. The temporal correlations were less consistent for the connectomes weighted by SIFT2 and 1-ODI. This may be an indication that connection weighting by FA and NDI produce more consistent and more reliable references for normal brain development in this period. 

Very few studies have assessed the asymmetry in brain's structural connectivity in the perinatal stage. Ratnarajah et al. observed differences in local and global efficiency between the left and right brain hemispheres in a population of 124 neonates between 36.9 and 42.7 gestational weeks \cite{ratnarajah2013structural}. However, their regression analysis did not reveal any association between age and local/global efficiency or betweenness centrality for a large set of brain structures considered. Our results, shown in Figure \ref{fig:Asymmetry}, are novel. They show a significant left laterality in the connections ending in the occipital and medial lobes and significant right laterality in the connections ending in the frontal and temporal lobes. Our results are consistent with previous researches on neonates, with rightward asymmetry for temporal lobes \cite{lehtola2019associations, hill2010surface, li2014mapping, li2015spatiotemporal}, rightward asymmetry for the frontal lobe \cite{vannucci2019cerebral}, and leftward asymmetry for occipital lobe \cite{gilmore2007regional, lehtola2019associations, vannucci2019cerebral}. Similar pattern is observed in adults with significant asymmetry in frontal, temporal and occipital lobes \cite{good2001voxel, toga2003mapping}. Remarkably, the main observations from all four weighting schemes are very similar.

\section{conclusion}

This work has proposed a novel computational framework for accurate quantitative assessment of the development of brain's structural connectome in the perinatal stage based on structural and diffusion MRI. The new framework relies on accurate alignment of white matter structures across many subjects using tensor- and FOD-based registration. This approach makes it possible to reduce the inter-subject variability and to reconstruct the developmental trajectories of the normal brain. Our experimental results, on 166 neonates between 33 and 44 postmenstrual weeks, show that the proposed framework can unveil relationships between several critical measures of brain connectivity and PMA. Connectome edge weighting based on FA and NDI are especially effective in uncovering strong trends in the structural connectivity measures. Our results show significant increases in network integration and segregation in the peri-natal stage. They also portray significant changes in connection strength and asymmetry between many nodes and lobes within and across brain hemispheres. The trends reconstructed in this work are much stronger and more consistent than the results reported in prior works on peri-natal strcutural brain connectivity. The normative developmental trends that have been discovered in this work can be used as reference baselines for comparing and contrasting normal and abnormal brain development in future works. Future works may also extend the proposed framework to analyzing brain connectivity in longitudinal and population studies.

\section{acknowledgements}
This research was supported in part by the National Institute of Neurological Disorders and Stroke, and Eunice Kennedy Shriver National Institute of Child Health and Human Development, and the National Institute of Biomedical Imaging and Bioengineering of the National Institutes of Health (NIH) under award numbers R01HD110772, R01NS128281, R01NS106030, R01EB031849, R01EB032366, and R01HD109395; and in part by the National Science Foundation (NSF) under grant number 212306. This research was also partly supported by NVIDIA Corporation and utilized NVIDIA RTX A6000 and RTX A5000 GPUs. The content of this publication is solely the responsibility of the authors and does not necessarily represent the official views of the NIH, NSF, or NVIDIA.

\section{conflict of interest}
The authors have no conflicts of interest.

\printendnotes

\bibliography{Yihan_references}



\end{document}